\documentclass[aps,prb,amsmath,amssymb,longbibliography,superscriptaddress,floatfix,showpacs,twocolumn]{revtex4-1}

\usepackage{graphicx}
\usepackage{mathdots}
\usepackage{epstopdf} 
\usepackage[usenames,dvipsnames]{color}
\usepackage{bm}

\usepackage{inputenc}
\usepackage{xcolor}
\usepackage{tikz}
\usepackage{physics}

\usepackage{booktabs}

\usepackage[colorlinks,bookmarks=false,citecolor=blue,linkcolor=red,urlcolor=blue]{hyperref}
\usepackage[colorlinks,bookmarks=false,citecolor=blue,linkcolor=red,urlcolor=blue]{hyperref}
\usepackage{xcolor}
\usepackage{mathdots}
\usepackage{float}
\usepackage{needspace}
\usepackage{soul}
\usepackage[usenames,dvipsnames]{color}

\begin{document}
\title{Break-even point of the quantum repetition code}

\author{\'Aron Rozgonyi}
\affiliation{Institute of Physics, 
E\"otv\"os University, 
H-1117 Budapest, Hungary}
\affiliation{Wigner Research Centre for Physics, H-1525 Budapest, Hungary}

\author{G\'abor Sz\'echenyi}
\affiliation{Institute of Physics, 
E\"otv\"os University, 
H-1117 Budapest, Hungary}
\affiliation{Wigner Research Centre for Physics, H-1525 Budapest, Hungary}

\begin{abstract}
Enhancing the lifetime of qubits with quantum code-based memories on different quantum hardware is a significant step towards fault-tolerant quantum computing. We theoretically show that the break-even point, i.e., preserving arbitrary quantum information  longer than the lifetime of a single idle qubit, can be beaten even with the quantum phase-flip repetition code in a dephasing-time-limited system. Applying circuit-based analytical calculation, we determine the efficiency of the phase-flip code as a quantum memory in the presence of relaxation, dephasing, and faulty quantum gates. Considering current platforms for quantum computing, we identify the gate error probabilities and optimal repetition number of quantum error correction cycles to reach the break-even point.

\end{abstract}

\maketitle

\section{Introduction}

\begin{figure}
\centering
\includegraphics[width=1.0\columnwidth]{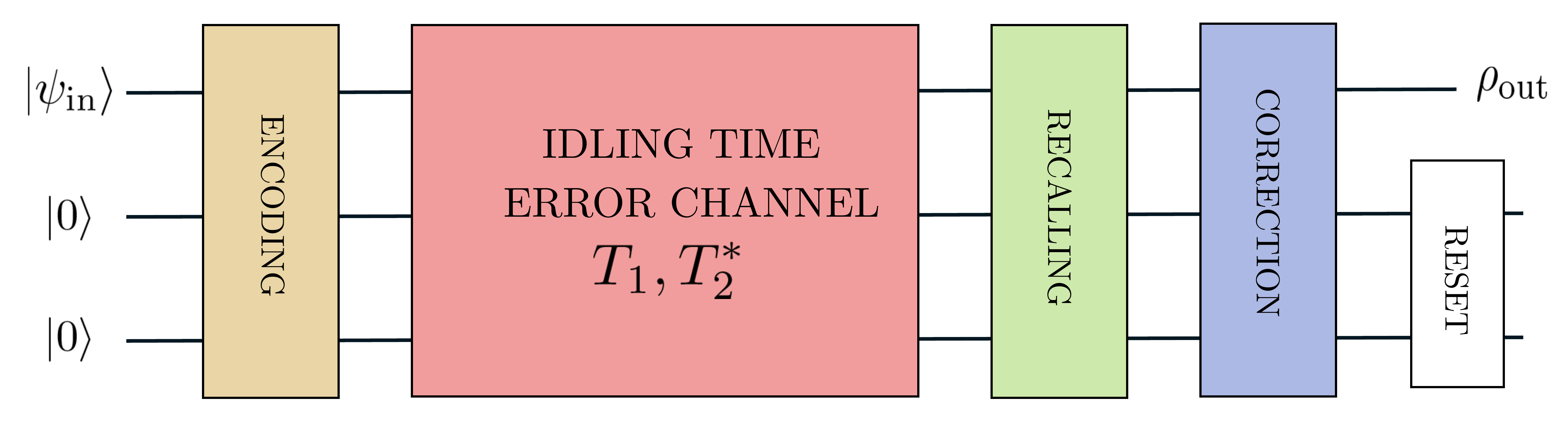}
\caption {\label{fig:schematic}
The schematic draw of a quantum code-based memory.  In the decoding step, the initial  state $\Psi_{in}$  is entangled  with a few ancillary qubits initialized in the ground state.  During the idling time, every qubit is affected by the noise sources, e.g., dephasing and relaxation. After the correction, the state is recalled  to one of the physical qubits. A reset of the ancilla qubits is necessary if we want to repeat this process. In many applications, the order of the recalling and correction processes is reversed. }
\end{figure}

Quantum memory, storing qubits for a sufficiently long time, is a key ingredient of almost any application in quantum communication \cite{Kimble2008}, computing \cite{doi:10.1126/science.1208517}, and sensing \cite{Zaiser2016}. Designing long-lived quantum memories is one of the recent challenges in the field of  quantum technology. It can be achieved by improving the lifetime  of each physical qubit and storing the logical quantum information in a collective quantum state of multiple physical qubits. The theory of quantum error correction proves  that  quantum codes are promising platforms for the fault-tolerant storage of quantum information. Typical quantum memory consists of the following processes:
encoding the quantum information into the quantum code, idling time, correction, and recalling the quantum state \cite{RevModPhys.87.307}. (See Fig. \ref{fig:schematic}.) Recalling is the inverse process of encoding when the logical information is refocused onto one physical qubit.

A large zoo of quantum codes is applicable for quantum memory \cite{QECCzoo}. However, 
only a few were demonstrated experimentally, such as  surface code, color code, Bacon-Shor code, and perfect code.
One of the most promising platforms is the surface code because, in theory, the logical errors could be reduced by scaling up the size of the system if the physical errors are below a threshold \cite{PhysRevA.86.032324, PhysRevA.83.020302}. Moreover, it can be realized on a 2D square grid of physical qubits with  connectivity only between the neighbors. However, because of initialization, gate, and measurement errors the \textit{break-even point} - i.e., preserving an arbitrary quantum state with surface code  longer than a single idle qubit's lifetime -  was not reached so far \cite{Acharya2022}. With other codes, such as cat code \cite{Ofek2016} and  a discrete-variable-encoded logical qubit \cite{Ni2023} the break-even point has been already reached.

The quantum repetition code is not a real quantum error correction code in the
sense that it cannot protect the logical information against any Pauli
error. Choosing any one of two types of repetition code (bit-flip code and phase-flip code) makes the quantum information resilient against Pauli X or Z errors, but at the same time, leaves it more vulnerable to the other error type \cite{9781107002173}. Nonetheless, mainly due to its simplicity, the repetition code is widely investigated
theoretically and used for benchmarking state-of-the-art quantum devices\cite{PhysRevA.97.052313, https://doi.org/10.48550/arxiv.2207.05568, Hicks_2022}. Up to now, the longest repetition code, containing 25 qubits, was implemented on a superconducting platform, and it was operated as a classical memory, where the noise was exponentially suppressed by scaling up the code \cite{Chen2021, Acharya2022}. However, in the case of repetition code-based quantum memories, any single-qubit  errors cannot be overcome by scaling up the system, only by combining the bit-flip and phase-flip quantum codes \cite{PhysRevA.52.R2493}. 

The simplicity of the repetition code comes from the fact that it requires only linear connectivity between the qubits, and for the implementation,  only three qubits are enough \cite{PhysRevX.8.021058, PhysRevA.102.022416}.  Thus, it was used in the first error correction experiments based on nuclear magnetic
resonance \cite{PhysRevLett.81.2152}, ion traps \cite{doi:10.1126/science.1203329}, superconducting circuits \cite{Reed2012, Kelly2015, Riste2015, Corcoles2015}, and NV-centers \cite{Waldherr2014}. Furthermore,
repetition code is the only error-correcting code that was realized on semiconductor qubits.  Phase-flip code-based quantum memory without mid-circuit measurements was implemented in Ge-based and in    Si-based  spin-qubit devices, but due to the imperfection of the gates, the lifetime of the logical qubit was  smaller than the lifetime of an idle physical qubit \cite{https://doi.org/10.48550/arxiv.2202.11530, Takeda2022}.

In this work, we aim to analytically determine the parameter regime (gate error, relaxation $T_1$, and dephasing $T_2^*$), where the repetition code-based quantum memory beats the break-even point. According to our  results, it can be achieved by applying phase-flip repetition code in a dephasing time-limited system ($T_1 > 2T_2^*$) if the gate errors are below a certain threshold. This result is a generalization of an earlier work, where only relaxation was taken into account \cite{PhysRevA.86.012333}.
We analytically study how the qubit's lifetime can be increased by repeating the quantum memory cycle sketched in Fig.~\ref{fig:schematic}. The investigated quantum circuit, shown in Fig. \ref{fig:circuit}, contains no mid-circuit measurements and it is composed of 1-qubit and controlled-NOT gates,  or it is composed of 1-qubit, controlled-Z, and controlled-S$^{-1}$ gates motivated by semiconducting hardware.

The rest of the paper is arranged as follows. In Sec.~\ref{repetition},  we briefly review the bit-flip and phase-flip repetition codes. The applied analytical methods based on density-matrix calculation are described in Sec.~\ref{method}. The break-even point of the repetition code is determined without gate errors in Sec.~\ref{withoutgateerror} and with gate errors in Sec.~\ref{gateerror}. The paper is enclosed with concluding remarks related to recent experiments in Sec.~\ref{discussion}.

\section{Repetition code} \label{repetition}

This chapter briefly introduces  the two kinds of three-qubit repetition codes, the bit-flip and the phase-flip repetition codes. In these, three physical qubits are used to encode the logical qubit, so that we can detect and correct one bit-flip or one phase-flip error, respectively. First, we discuss the bit-flip repetition code and then compare it with the phase-flip repetition code. 

In the case of the bit-flip repetition code, the logical information $|\Psi\rangle _\textrm{in} = \alpha |0\rangle +\beta |1\rangle$ is encoded as $|\Psi\rangle = \alpha |000\rangle +\beta |111\rangle$, where $|000\rangle$ ($ |111\rangle$) means the product-state of three qubits, all being in the ground state $|0\rangle$ (excited state $|1\rangle$). Bit-flips are detected by measuring the parity of the neighboring physical qubits. For this purpose, we perform projective measurement with the stabilizer operators $Z_1Z_2$ and $Z_2Z_3$, where Pauli- Z operator ${Z}_i$ is acting on the $i$-th physical qubit.  Applying a majority-vote decoder on measurement outcomes, called error syndromes, one bit-flip error is corrected by acting with an $X$ gate on the faulty qubit chosen by the decoder. After the correction, the qubit is returned to the codespace then, as a final step, the information is recalled from the encoded state.  

Two or more bit-flip errors cannot be corrected in the three-qubit repetition code. However, the logical state still becomes more robust to bit-flip errors. On the other hand, this code cannot detect phase-flip errors, and because of the  encoding,  more qubits are exposed to its harmful effects, thus in bit-flip code, the logical qubit is more vulnerable to phase-flip errors. 

Realizing projective measurement with the stabilizers requires two extra ancilla qubits. Moreover, from an experimental point of view, measurements are the most time-consuming and faulty steps in quantum error correction schemes. Therefore, these mid-circuit measurements are sometimes replaced by a three-qubit Toffoli gate. Toffoli gates have already been realized by resonant driving \cite{PhysRevB.100.085419,Takeda2022}, but even in some recent experiments, it has been composed of single-qubit and two-qubit gates \cite{https://doi.org/10.48550/arxiv.2202.11530}. 

In the phase flip repetition code, the initial state $\Psi_\textrm{in}$ is encoded as $|\Psi\rangle = \alpha |---\rangle +\beta |+++\rangle$, where $|\pm\rangle=\frac{1}{\sqrt{2}}(|0\rangle\pm|1\rangle)$. Compared with bit-flip code, the role of Pauli-X and Pauli-Z operators are exchanged, then in phase-flip repetition code, we can correct one phase-flip ($Z$ error) by measuring with the stabilizer $X_1X_2$ and $X_2X_3$. The bit-flip code is converted to phase-flip code by acting on every qubit with Hadamard gates before and after transmission through the error channel.

The quantum circuit of the repetition code-based quantum memory is realized in the following way: (1) initially, one qubit holds the information, while the two other ancilla qubits are in their ground state; (2) for the encoding  two CNOT gates are applied; (3) only in case of phase-flip code Hadamard-gates are acting on every qubit; (4) errors are acting during the idling time; (5) only in case of phase-flip code Hadamard-gates are acting on every qubit; (6) two other CNOT gates for the recalling; (7) Toffoli gate for the error correction. 

\begin{figure}
\centering
\includegraphics[width=1.0\columnwidth]{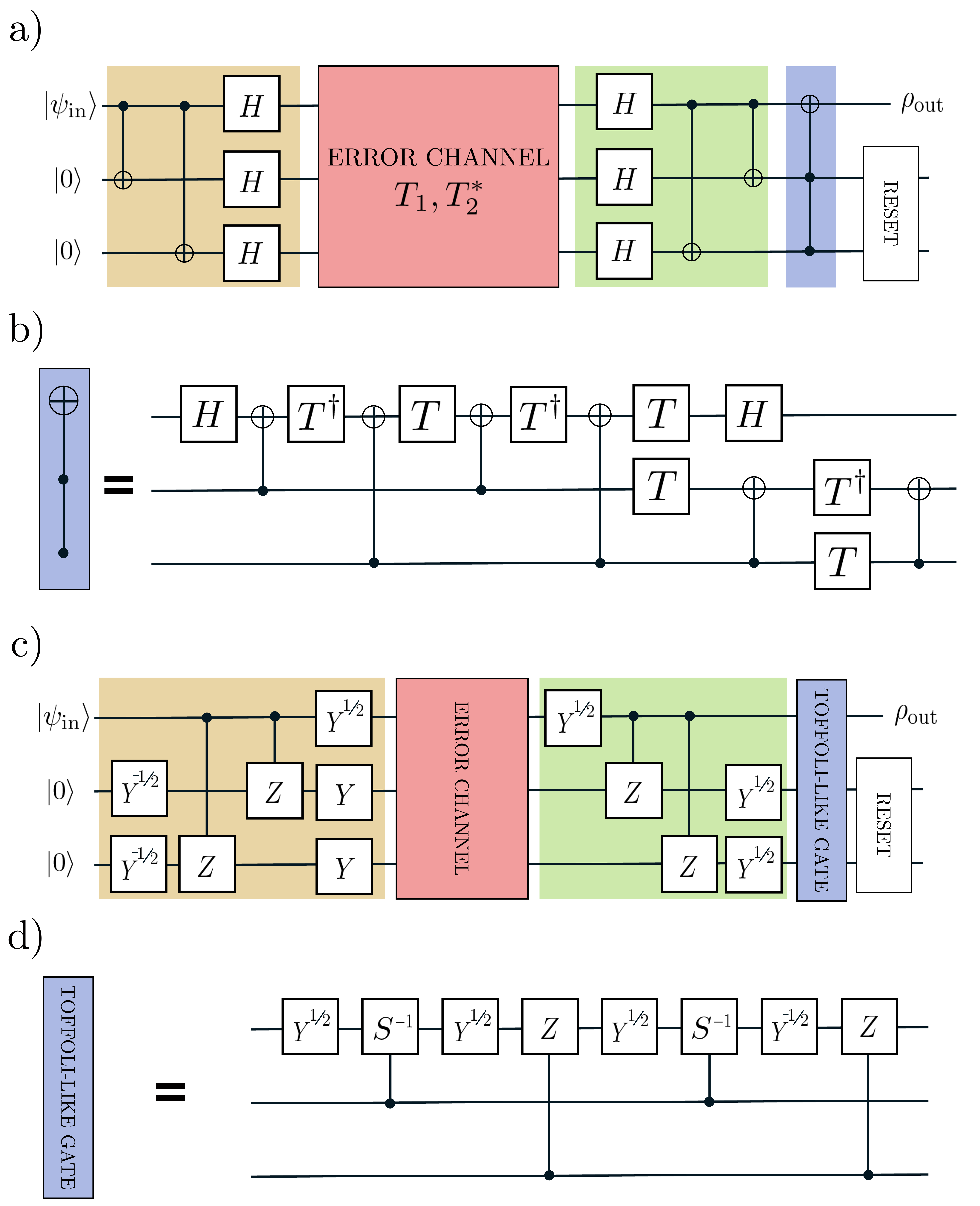}
\caption{\label{fig:circuit}
Quantum circuit of the phase-flip repetition code. (a) and (b), it is decomposed of 1-qubit
gates and CNOT gates.  (c) and (d), it is decomposed of 1-qubit
gates, controlled-Z, and controlled-S$^{-1}$ gates}
\end{figure}

The circuit diagram of the phase-flip repetition code is plotted in Fig. \ref{fig:circuit}a and b, composed of 1-qubit gates and CNOT gates. This decomposition is motivated by experimental platforms, where CNOT gates are easy to implement between any two of the three physical qubits used for the repetition code. For the bit-flip code, one should omit the Hadamard gates in Fig. \ref{fig:circuit}a.
In Fig. \ref{fig:circuit}c and d, we show the experimental realization of phase flip code composed of 1-qubit, controlled-Z, and controlled-S$^{-1}$ gates recently used on semiconducting hardware \cite{https://doi.org/10.48550/arxiv.2202.11530}. In our  analytical calculation, we will use these two different decompositions of the phase-flip repetition code.

\section{Method: density-matrix calculation} \label{method}

Density-matrix description is a useful  tool for handling small quantum systems, such as 
repetition codes \cite{PhysRevA.95.012306, Baek_2020}, because of the following advantages. (i) We can easily take into account any noise channel described by Kraus operators without any simplification, e.g., Pauli-twirling approximation \cite{PhysRevA.90.062320, PhysRevA.88.012314}. (ii) Instead of performing projective measurements and sampling the outcome at the end of each quantum error correction cycle, we derive the density matrix and directly extract the probability density of the measurement from it.  (iii) This simulation method opens up the possibility of deriving analytical formulas, for example, for the fidelity of repetition-code-based quantum memories.

This chapter shows how to build the density matrix calculation for the quantum error correction circuits shown in Fig. \ref{fig:circuit}.  We consider two kinds of errors: independent qubit errors during the idling time  and faulty gates. We neglect other potential noise sources. Thus, for example, we assume perfect initialization and reset. 

During the idling time, $t_\textrm{idle}$ the qubits are coupled to the environment, which is usually modeled  by amplitude and phase damping channels characterized by the relaxation time $T_1$ and  dephasing time $T_2^*$, respectively. These noisy single-qubit quantum channels are described by a density-matrix evolution: 
\begin{equation}
\rho \rightarrow \sum_{k=1}^3 E_k\rho E_k^\dagger,
    \end{equation}
where $E_k$ Kraus-operators are chosen in the following way \cite{PhysRevA.86.062318}:
\begin{eqnarray}
  E_1&= &\begin{pmatrix}
\sqrt{1-\alpha} & \\
0 & \sqrt{1-\gamma}
\end{pmatrix}, 
\; 
    E_2= \begin{pmatrix}
0 & \sqrt{\gamma}\\
0& 0
\end{pmatrix}, \nonumber
\\
E_3&=& \begin{pmatrix}
\sqrt{\alpha} & 0\\
0 & 0
\end{pmatrix}.
\end{eqnarray}
The parameters in the Kraus operators have the form
$\gamma=1-e^{-t_\textrm{idle}/T_1}$, $\alpha=1-e^{-2t_\textrm{idle}/T_2^*}$. Going through the error channel the density matrix of a single qubit is transformed as
\begin{equation}
\begin{pmatrix}
\rho_{00} & \rho_{01}\\
\rho_{10} & \rho_{11}
\end{pmatrix}
\rightarrow
    \begin{pmatrix}
1-e^{-t_\textrm{idle}/T_1}\rho_{11} & e^{-t_\textrm{idle}/T_2}\rho_{01}\\
e^{-t_\textrm{idle}/T_2}\rho_{01}^* & e^{-t_\textrm{idle}/T_1}\rho_{11}
\end{pmatrix},
\end{equation}
where 
\begin{equation} \label{T2def}
\frac{1}{T_2}=\frac{1}{T_2^*} + \frac{1}{2T_1}.
\end{equation}

Pure dephasing ($T_1=0$, $T_2^*=\textrm{final}$) is an asymmetric depolarization channel where the qubit suffers from a simple Pauli-Z error. However, relaxation, where  amplitude damping always goes together with dephasing, cannot be exactly captured by the asymmetric depolarization channel. Therefore, it is usually approximated by the Pauli twirling, with which the relaxation is decomposed into the sum of Pauli X, Y and Z channels \cite{PhysRevA.86.062318}. In our work, we use the exact time evolution instead of the Pauli twirling approximation.

The error channel of the faulty gates is taken into account by the depolarizing error model, which is parameterized by the probability $p_1$ for the single-qubit gates and $p_2$ for the two-qubit gates. In the case of a single qubit depolarizing channel, after  every single-qubit gate the density matrix is mapped as
\begin{equation}
\rho \rightarrow \left(1-\frac{4p_1}{3}\right)\rho + \frac{2p_1}{3} I_2,
    \end{equation}
where $I_2$ is a $2\times 2$ identity matrix. The Kraus decomposition of this transformation is given by the following four Kraus operators, $\left\{\sqrt{1-p_1}I_2\right.$, $\sqrt{\frac{p_1}{3}}X$, $\sqrt{\frac{p_1}{3}}Y$, $\left.\sqrt{\frac{p_1}{3}}Z\right\}$. Two-qubit faulty gates are defined by similar  transformations after every two-qubit gates 
\begin{equation}
\rho \rightarrow \left(1-\frac{16p_2}{15}\right)\rho + \frac{4p_2}{15} I_4.
    \end{equation}
The Kraus decomposition of this two-qubit channel is given by  sixteen Kraus operators, $\left\{\sqrt{1-p_2}I_4\right.$, $\sqrt{\frac{p_2}{15}}I_2X$, $\sqrt{\frac{p_2}{15}}I_2Y$, $\sqrt{\frac{p_2}{15}}I_2Z$, $\sqrt{\frac{p_2}{15}}XX$, $\left.\dots, \sqrt{\frac{p_2}{15}}ZZ\right\}$, where the tensor products of the Pauli-matrices $\left\{I_2,X,Y,Z\right\}$ have  appeared.  The first (second) Pauli-matrix in the Kraus-operator is acting on the first (second) output of the two-qubit quantum gate. No Pauli-matrix is acting on the qubit, which was not involved in the two-qubit operation. 

The use of the depolarizing channel for  faulty gates is a simplifying assumption, but it is widely used in simulations and theoretical analyses because it keeps low the number of
parameters in the calculation. 

Here, the quantum error correction cycle operates as a quantum memory, therefore, the longest step in Fig. \ref{fig:schematic} is the idling time $t_\textrm{idle}$ between the encoding and the correction.  Based on this, gate operation times are neglected, and the length of the quantum error correction cycle is assumed to be $t_\textrm{idle}$. If we repeat the error correction cycles $N$ times, then the total storage time is $t_\textrm{tot}=Nt_\textrm{idle}$. 
We will investigate this repetition   with and without the reset of the ancilla qubits. In the case of the former one, after the recalling and correction, the two ancilla qubits are instantaneously set to their ground state, which can experimentally be achieved by using two new, already relaxed ancilla qubits. If there is no  reset of the ancilla qubits, then the outcome of the error correction cycle will be the three-qubit input state of the next cycle. In our scheme, encoding, and recalling must be done in every round. 

At the end of the quantum information storage procedure, we trace over the degrees of freedom of the ancilla qubits and obtain a $2\times2$ density matrix $\rho_\textrm{out}$. The efficiency of the quantum memory is characterized by the fidelity, which measures the closeness of the initial state $\Psi_\textrm{in}$ and the output density matrix $\rho_\textrm{out}$ in the following way, $F=\langle\Psi_\textrm{in}|\rho_\textrm{out}|\Psi_\textrm{in}\rangle $. It depends on the initial state, and we want to store arbitrary quantum states. Therefore the fidelity is integrated over the Bloch sphere to calculate the average of the fidelity
\begin{equation}
\mathcal{F}=\int_\textrm{Bloch sphere} \langle\Psi_\textrm{in}|\rho_\textrm{out}|\Psi_\textrm{in}\rangle \;d\alpha\;d\beta.
\end{equation}
The average fidelity $\mathcal{F}$ is a simple quantity to characterize the accuracy of the quantum storage, hence
the main goal of this paper is to calculate it for different quantum code-based memories as a function of relaxation and dephasing time, gate errors, idling time, and the number of error correction cycles.

\section{Break-even point - without gate error} \label{withoutgateerror}

To determine the parameter range required for reaching the break-even point, we need to compare the accuracy of a quantum code-based memory and a single idle qubit. Hence, first of all, we calculate  the average fidelity of a single qubit suffering from decoherence and relaxation for a time $t_\textrm{idle}$,
\begin{equation} \label{eq:f_idle}
 \mathcal{F}_\textrm{idle} = \frac{1}{6}\left(3 +e^{-\frac{t_\textrm{idle}}{T_1}}+2e^{-\frac{t_\textrm{idle}}{T_2}} \right).
\end{equation}
This is the same formula as Eq. (1) in Ref. \onlinecite{OBrien2017}. In this chapter, the main goal is to calculate the average fidelity without  gate errors ($p_1=p_2=0$) for a bit-flip and a phase-flip repetition code-based quantum memory, $\mathcal{F}_\textrm{bit}$ and $\mathcal{F}_\textrm{phase}$, respectively, and compare these values with $\mathcal{F}_\textrm{idle}$. The definition of beating the break-even point is that the condition $\mathcal{F}_\textrm{bit/phase}>\mathcal{F}_\textrm{idle}$ is fulfilled for a certain storage time.

After one cycle of the error correction in a three-qubit bit-flip repetition code model, the average fidelity is given by the formula
\begin{equation}
\mathcal{F}_\textrm{bit}=\frac{1}{6}\left(3 +2 e^{-\frac{3t_\textrm{idle}}{T_2}}-2e^{-\frac{3t_\textrm{idle}}{T_1}} + 3e^{-\frac{2t_\textrm{idle}}{T_1}}\right).
\end{equation}
For arbitrary values of storage, relaxation, and dephasing time, $\mathcal{F}_\textrm{bit}$ is always smaller or equal to $\mathcal{F}_\textrm{idle}$, hence the break-even point cannot be reached by the bit-flip code. The qualitative understating of this statement is the following, a bit-flip code cannot correct pure dephasing (Pauli-Z) error. Relaxation is approximately a combination of Pauli-X, Y and Z errors, and only the first of them, the Pauli-X error is mitigated, meanwhile, the information becomes more vulnerable to the Pauli-Z error.  As a consequence, the lifetime of a qubit cannot be enhanced by a bit-flip repetition code, therefore further on, we focus on only the phase-flip repetition codes. 

\begin{figure}
\centering
\includegraphics[width=1.0\columnwidth]{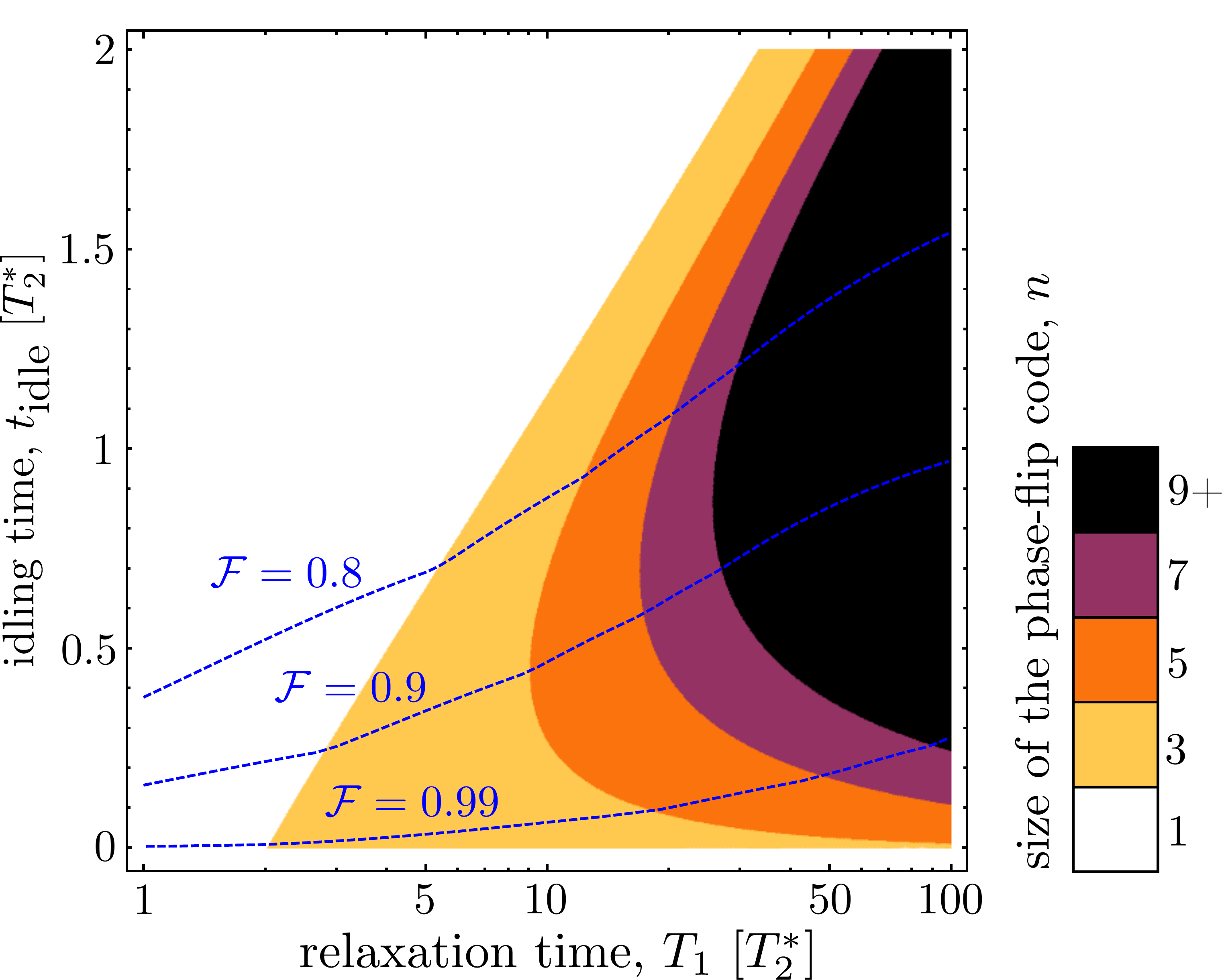}
\caption{\label{fig:noerror}The size of the phase-flip repetition code giving the largest fidelity after one cycle of  error correction as well as three iso-fidelity contours as a function of relaxation and idling times in units of $T_2^*$. Gate errors are neglected.  For $n=1$ (white region) the idle qubit gives better fidelity than a quantum code. In a  dephasing-time limited system  $(T_1 > 2T_2^*)$ if the idling time is shorter than a threshold value (colorful region) then the break-even point is beaten. 
}
\end{figure}

After one cycle of  3-qubit phase flip code-based quantum error correction, the average fidelity is decreased as
\begin{eqnarray}\label{eq:f_phase}
\mathcal{F}_\textrm{phase}&=&\frac{1}{12}\left[6 +2 e^{-\frac{3t_\textrm{idle}}{T_1}}-2e^{-\frac{3t_\textrm{idle}}{T_2}}\nonumber\right.
\\ &+& \left.3e^{-\frac{t_\textrm{idle}}{T_2}}\left(1+e^{-\frac{2t_\textrm{idle}}{T_1}}\right)\right].
\end{eqnarray}
This value is larger than $\mathcal{F}_\textrm{idle}$, if $T_1 > 2T_2^*$ and the storage time $t_\textrm{idle}$ is below a threshold value. This statement is visualized in Fig. \ref{fig:noerror}, the white region shows the parameter range, where the single qubit performs better $\mathcal{F}_\textrm{idle}>\mathcal{F}_\textrm{phase}$, while in the colorful region (yellow-orange-purple-black) the quantum code  performs better $\mathcal{F}_\textrm{phase}>\mathcal{F}_\textrm{idle}$.

The phase-flip repetition code mitigates (amplifies) the fidelity reduction due to the pure dephasing (relaxation), therefore, the break-even point is beaten only in a dephasing-time limited system, where $T_1 > 2T_2^*$.

In the rest of the paper,  we focus on the 3-qubit repetition codes, but now in this single paragraph, we investigate longer ones. In an $n$-qubit phase-flip code, where $n$ is an odd number, the logical information is encoded via the superposition of collective states of $n$ qubits
, all being in the state $|+\rangle$ or all being in the state $|-\rangle$. After the measurement with the stabilizer operators, which are products of two Pauli-X operators acting on the neighboring qubits, we can detect and also correct $(n-1)/2$ phase-flip errors. Calculation of the average fidelity after one cycle of the error correction using $n$-qubit phase-flip repetition code  is analytically tractable and leads to the following
\begin{widetext}
\begin{equation} \label{eq:f_general}
\mathcal{F}_\textrm{phase,$n$}=\frac{1}{3} + \frac{1}{3}\frac{1}{2^n} \sum_{k=0}^{\frac{n-1}{2}} \binom{n}{k} \left[\left(1- e^{-\frac{t_\textrm{idle}}{T_2}}\right)^k \left(1+ e^{-\frac{t_\textrm{idle}}{T_2}}\right)^{n-k}\right.+\left. \left(e^{-\frac{t_\textrm{idle}}{T_1}}- e^{-\frac{t_\textrm{idle}}{T_2}}\right)^k \left(e^{-\frac{t_\textrm{idle}}{T_1}}+ e^{-\frac{t_\textrm{idle}}{T_2}}\right)^{n-k}\right].
\end{equation}
\end{widetext}
Substituting $n=1$ and $n=3$ into this formula, we get back Eq. (\ref{eq:f_idle}) and Eq. (\ref{eq:f_phase}), respectively. 
The dephasing error can be overcome  by scaling up the system, but the relaxation error cannot, in fact, relaxation causes an increasingly harmful effect.  Therefore, a finite-size phase-flip code is an optimal choice if we would like to maximize the fidelity. 
According to Eq. (\ref{eq:f_general}), in Fig. \ref{fig:noerror} we plot the length of the optimal  phase-flip code as well as three iso-fidelity contours as a function of relaxation and idling times.  It is worth choosing phase-flip codes longer than three only if $T_1\gg T_2^*$. 

If we would like to store the quantum information for a time $t_\textrm{tot}$, then we can execute one error correction cycle as it was discussed so far in this chapter, or we can repeat the cycles $N$ times one after another. Assuming uniform time step, every cycle has a duration $t_\textrm{tot}/N$. Repetition is executed with or without the reset of the ancilla qubits between the error correction cycles.  In the former case, the values  of two ancilla qubits are instantaneously reset to their original ground state. The average fidelity after $N$ rounds of the error correction reads 
\begin{eqnarray}\label{eq:f_phase_reset}
\mathcal{F}_\textrm{phase}^\textrm{reset}(N)&=&\frac{1}{6}\left\{3 + e^{-\frac{3t_\textrm{tot}}{T_1}}+\frac{1}{2^N}e^{-\frac{3t_\textrm{tot}}{T_2}}
\left[\left(3 e^{\frac{2t_\textrm{tot}}{NT_2}}-1\right)^N\right.\right.\nonumber
\\ &+&\left.\left.\left(3 e^{\frac{2t_\textrm{tot}}{NT_2}}e^{\frac{-2t_\textrm{tot}}{NT_1}}-1\right)^N \right]\right\}.
\end{eqnarray}
It can be proven that $\mathcal{F}_\textrm{phase}^\textrm{reset}(N)$ is a monotonically increasing function of $N$. 
Therefore, in this simple model, where gate errors are neglected, the  lifetime of the logical qubit in a phase-flip code-based quantum memory can be enhanced by increasing the number of error correction cycles. If we repeat the cycles infinitely many times, the errors due to the pure dephasing disappear, however, the fidelity 
\begin{equation}
\lim_{N \to \infty} \mathcal{F}_\textrm{phase}^\textrm{reset}(N) = \frac{1}{3}\left(2+e^{-\frac{3t_\textrm{tot}}{T_1}}\right)
\end{equation}
is decreasing in time because of the relaxation. 

In the other case, we repeat the cycles without resetting the ancilla qubits. It means, that the 3-qubit outgoing state of an error correction cycle will be the initial state of the next one. Only for clarification, we note, that the ancilla qubits are not measured  in our scheme.  The average fidelity after two cycles of error correction with total time $t_\textrm{tot}$ reads
\begin{eqnarray}
\mathcal{F}_\textrm{phase}^{\textrm{no-reset}} (N=2)&=&\frac{1}{24}\left[12 +4 e^{-\frac{3t_\textrm{tot}}{T_1}}-2e^{-\frac{3t_\textrm{tot}}{T_2}}\nonumber\right.
\\ &+& \left.5e^{-\frac{t_\textrm{tot}}{T_2}}\left(1+e^{-\frac{2t_\textrm{tot}}{T_1}}\right)\right],
\end{eqnarray}
and it is always smaller than Eq. (\ref{eq:f_phase}), which is the average fidelity after one cycle with the same storage time. This result implies, that even without gate errors, it is not worthwhile to  repeat the error correction cycles if we are not capable of resetting the ancilla qubits.

\section{Break-even point - with gate error} \label{gateerror}

In this chapter, we calculate the average fidelity in the presence of gate errors. These results depend on the gate decomposition of the error correction cycles, therefore, we choose two special ones which are investigated further on. These two quantum circuits, the CNOT-based and the CS-CZ-based have already been introduced in Fig.  \ref{fig:circuit}. 

We assume that the gate errors are small, $\left\{p_1,p_2\right\}\ll1$, therefore 
the drop in the fidelity
 is linearized in the variables of $p_1$ and $p_2$. The average fidelity after one error correction cycle reads as
\begin{equation} \label{eq:F_gateerror}
\mathcal{F}_\textrm{phase}^\textrm{error}=\mathcal{F}_\textrm{phase} - p_1 f_1 - p_2 f_2,
\end{equation}
where the fidelity without gate errors $\mathcal{F}_\textrm{phase}$ is formulated in Eq. (\ref{eq:f_phase}), furthermore, $f_1$ is slightly different for the CNOT-based quantum circuit
\begin{eqnarray}
f_1^\textrm{CNOT}&=&\frac{4}{3}\left(2e^{-\frac{-3t_\textrm{idle}}{T_1}}-2e^{-\frac{-3t_\textrm{idle}}{T_2}}+2e^{-\frac{-t_\textrm{idle}}{T_2}}\right.\nonumber\\&+&\left.3e^{-t_\textrm{idle}\left(\frac{1}{T_2}+ \frac{2}{T_1}\right)} \right)
\end{eqnarray}
and for the CS-CZ-based quantum circuit
\begin{eqnarray}
\label{eq:F_gateerrorCZ1}
f_1^\textrm{CS-CZ}&=&\frac{4}{9}\left(4e^{-\frac{-3t_\textrm{idle}}{T_1}}-6e^{-\frac{-3t_\textrm{idle}}{T_2}}+5e^{-\frac{-t_\textrm{idle}}{T_2}}\right.\nonumber\\&+&\left.7e^{-t_\textrm{idle}\left(\frac{1}{T_2}+ \frac{2}{T_1}\right)} \right).
\end{eqnarray}
Surprisingly, the prefactor of the two-qubit gate error 
\begin{eqnarray}
\label{eq:F_gateerrorCZ2}
f_2&=&\frac{8}{45}\left(8e^{-\frac{-3t_\textrm{idle}}{T_1}}-8e^{-\frac{-3t_\textrm{idle}}{T_2}}+11e^{-\frac{-t_\textrm{idle}}{T_2}}\right.\nonumber\\&+&\left.12e^{-t_\textrm{idle}\left(\frac{1}{T_2}+ \frac{2}{T_1}\right)} \right)
\end{eqnarray}
is the same for the two realizations. If the relaxation and dephasing errors are negligible, the fidelity is dropped  by $184p_2/45$ due to the faulty two-qubit gates and dropped  by  $10p_1/3$ or $40p_1/9$   due to the faulty single-qubit gates in the case of CNOT-based or in the case of CS-CZ-based quantum circuit, respectively. The gate errors ($p_1$, $p_2$) required to reach the break-even point can be determined for any parameter point of the colored region in Fig. \ref{fig:noerror} by fulfilling the inequality $\mathcal{F}_\textrm{phase}^\textrm{error}>\mathcal{F}_\textrm{idle}$.

\begin{figure}
\centering
\includegraphics[width=1.0\columnwidth]{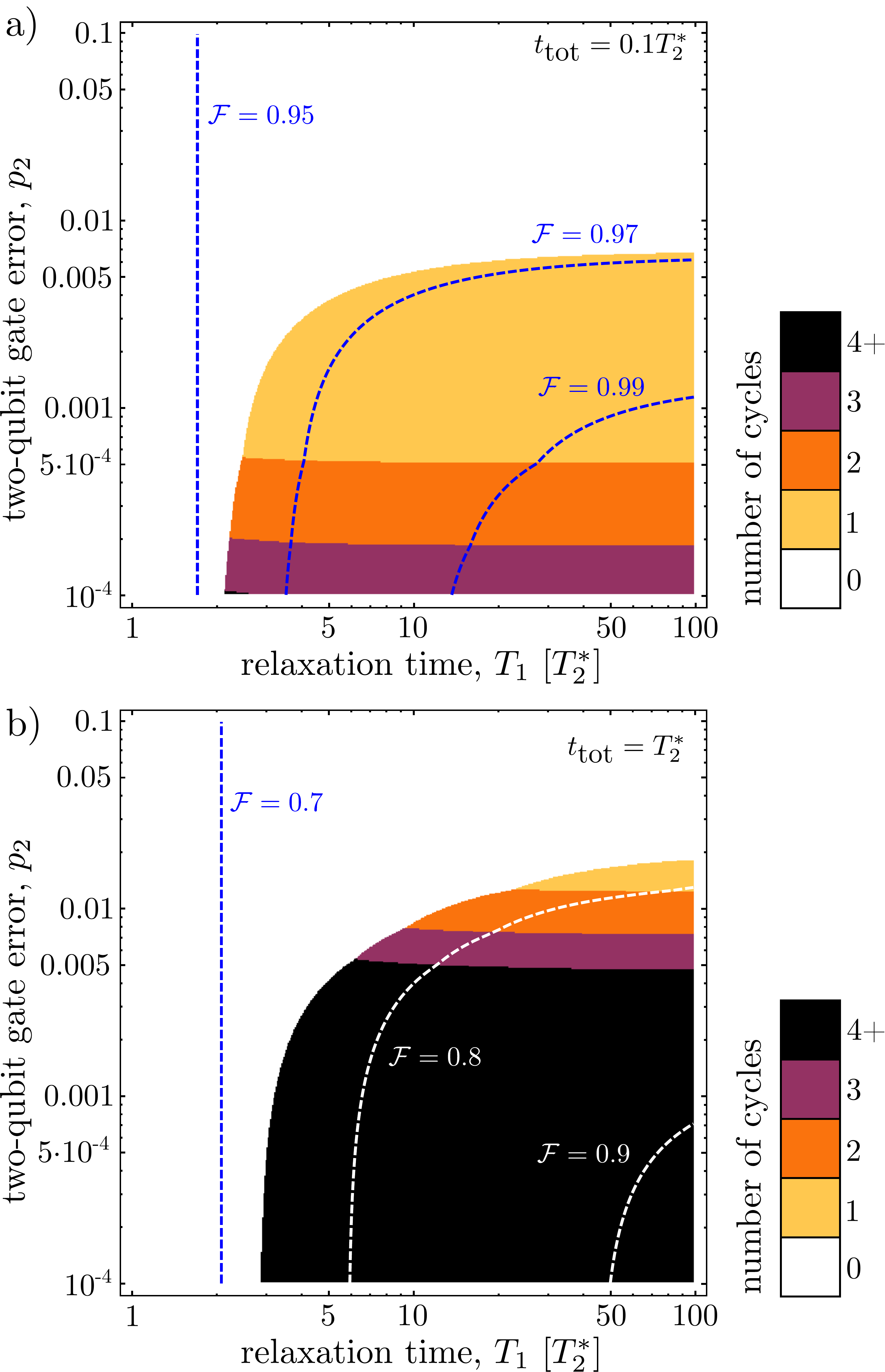}
\caption {\label{fig:gate_error}
The optimal number of the error correction cycles in a phase-flip code-based quantum memory as well as three iso-fidelity contours as a function of the relaxation time and the two-qubit gate error. Single-qubit gate errors are neglected $p_1=0$.   For $N=0$  (white region) the idle qubit gives better fidelity than a quantum code. The colored region (yellow-orange-purple-black) shows the parameter range where  the break-even point is reached. Black color means that the optimal number of repetitions is four or more. In Fig. a) the storage time is $t_\textrm{tot}=0.1T_2^*$ , while in Fig. b) it is $t_\textrm{tot}=T_2^*$.}
\end{figure}

It was shown in the previous section, that in the case of perfect gates and reset, the fidelity can be maximized by repeating the error correction cycles as often as possible. These changes due to gate errors, and a finite number of repetitions is the optimal choice, which is demonstrated in Fig. \ref{fig:gate_error}, where the number of error correction cycles for the maximal fidelity  as well as three iso-fidelity contours are plotted as the function of the relaxation time and the two-qubit gate errors for storage time $t_\textrm{tot}=0.1T_2^*$ in figure a, and $t_\textrm{tot}=T_2^*$ in figure b. In the experiments, typically the two-qubit gate error exceeds the single-qubit gate error, therefore, in these plots we assume perfect single-qubit gates $p_1=0$. The parameter range where the single idle qubit is an optimal choice against the phase-flip code-based quantum memory is colored white. In the colorful region, the break-even point is beaten, and it is worthwhile to repeat cycles more often if the two-qubit gate errors can be reduced.

\section{Discussion} \label{discussion}

In this chapter, we discuss the relevance of the above-derived analytical results in light of current experiments.  
Recently developed quantum processing units with the largest qubit numbers utilize superconducting architectures, and therefore, first we discuss the performance of the phase-flip code-based quantum memory on these platforms. In Table \ref{tab:super} we collect the mean value of relaxation and decoherence times, and we calculate the pure dephasing times from them according to Eq. (\ref{T2def}) for a few superconducting quantum processors. In these systems, the qubit lifetime is limited by the relaxation because the $T_1$ relaxation time is shorter or approximately the same as the pure dephasing time $T_2^*$. Except for the ibm\_jakarta, all the other quantum chips do not fit the condition $T_1 > 2T_2^*$  required to reach the break-even point. In the case of the ibm\_jakarta, if the waiting time $t_\textrm{idle}$ is smaller than 11 $\mu s$ and the gates are perfect, then $\mathcal{F}_\textrm{idle}<\mathcal{F}_\textrm{phase}$. However, due to the imperfect operations, the mean CNOT error is around $8\cdot10^{-3}$ for  ibm\_jakarta \cite{ibmq2023}, hence we cannot demonstrate the break-even point. In conclusion, these superconducting quantum chips are unsuitable platforms for implementing an efficient phase-flip code-based quantum memory.

\begin{center}
\begin{table}[h]
\begin{tabular}{ |c||c|c|c|  }
 \hline
 \multicolumn{4}{|c|}{Superconducting architectures} \\
 \hline
quantum chip & $T_1$ [$\mu$s] &$T_2$ [$\mu$s]& $T_2^*$ [$\mu$s]\\
 \hline
Google Sycamore  \cite{Acharya2022} & 20    &30&   120\\
IBM ibm\_washington  \cite{ibmq2023} & 96     &81 &   140 \\
IBM ibm\_montreal \cite{ibmq2023} & 101    &64 &   94  \\
IBM ibm\_nairobi\cite{ibmq2023} &   114  & 76    & 114 \\
 IBM ibm\_jakarta \cite{ibmq2023} &125 & 38 &  45  \\
 Rigetti Aspen M-2   \cite{rigetti2023}&33 & 19  &  27 \\
 Rigetti Aspen M-3 \cite{rigetti2023}  &22 & 24  &  53 \\
 \hline
\end{tabular}
\caption{\label{tab:super}The mean value of the relaxation, decoherence and dephasing times for different superconducting quantum computers. }
\end{table}
\end{center}

The lifetime of qubits in semiconductor quantum devices is typically limited by the dephasing, and not by the relaxation, i.e., $T_1\gg T_2^*$ as shown in Table \ref{tab:semi}.  Dephasing time is mitigated by a Hahn-echo experiment, which technique is also compatible with the quantum memory scheme. In this case, an inverse pulse is applied in the middle of the idling time. Even after the Hahn-echo experiment $T_1\gg T_2^\textrm{Hahn}$, therefore semiconductor quantum devices are promising platforms to demonstrate the efficiency of the phase-flip code-based quantum memory.

\begin{center}
\begin{table}[h]
\begin{tabular}{ |c||c|c|c|  }
 \hline
 \multicolumn{4}{|c|}{Semiconductor devices} \\
 \hline
platform & $T_1 [\textrm{ms}]$  &$T_2^*$ [$\mu$s] & $T_2^\textrm{Hahn}$[$\mu$s] \\
 \hline

  3-qubit Si \cite{Takeda2022}  &22 & 1.8  &  43 \\
 \hline
  4-qubit Ge \cite{Hendrickx2021,https://doi.org/10.48550/arxiv.2202.11530}  &1-16 & 0.15-0.4  &  3-5 \\
 \hline
   2-qubit SiGe \cite{doi:10.1126/sciadv.abn5130}  &24-48 & 1.7-2.3  &  23-100 \\
 \hline
  6-qubit Si \cite{Philips2022}  &$\gg$ 1 & 3-5  &  14-27 \\
 \hline

\end{tabular}
\caption{\label{tab:semi}The relaxation, pure dephasing and spin-echo dephasing times for a few semiconductor quantum devices. Phase-flip code has already been implemented on the 3-qubit Si  and 4-qubit Ge  devices.}
\end{table}
\end{center}

After substituting Eqs. (\ref{eq:f_idle}), (\ref{eq:f_phase}), (\ref{eq:F_gateerror}), (\ref{eq:F_gateerrorCZ1}), and (\ref{eq:F_gateerrorCZ2}) into the inequality $\mathcal{F}_\textrm{idle}<\mathcal{F}_\textrm{phase}^\textrm{error}$ we  define the range of  the single- and two-qubit gate errors for which the break-even point is reached. For example, after an idling time $t_\textrm{idle}=0.1T_2^\textrm{Hahn}$, the fidelity of a single idle qubit is dropped to 0.97. It is enhanced by one cycle of error correction with  phase-flip code if the single- and two-qubit gate errors satisfy the condition 
\begin{equation}
p_2<0.0067-1.1p_1.
\end{equation}
Similarly,  after an idling time $t_\textrm{idle}=T_2^\textrm{Hahn}$, which is a time 40 $\mu$s and 4 $\mu$s in experiments [\onlinecite{Takeda2022}] and  [\onlinecite{https://doi.org/10.48550/arxiv.2202.11530}], respectively, the fidelity of a single idle qubit is dropped to 0.79. It is enhanced by one cycle of error correction with  phase-flip code if 
\begin{equation}
p_2<0.019-1.3p_1. \label{eq:gate_result}
\end{equation}
We should note that these calculations are valid for the experiments [\onlinecite{https://doi.org/10.48550/arxiv.2202.11530}], but give only a rough estimation for the experiment [\onlinecite{Takeda2022}], because there the Toffoli gate was implemented by resonant driving and not by the decomposition into single- and two-qubit gates.  It is also worth noticing that our analytical results are accurate if the idling time is much larger than the time duration of the gate sequence, which is around 0.5-1 $\mu$s in both experiments.  According to Eq. (\ref{eq:gate_result}),   gate errors of around 1-2\% are required to demonstrate the break-even point with a phase-flip code. 

\section{Summary}

To summarize, we collect all the necessary ingredients to hit the break-even point with a repetition code-based quantum memory.
\begin{itemize}
    \item{Apply phase-flip repetition code.}
    \item{Apply on a dephasing time-limited platform, where $T_1 > 2T_2^*$.}
    \item{The idling time should be below a threshold value, as it is shown in Fig. \ref{fig:noerror}.}
    \item{The gate errors should be below a threshold value, but the exact condition depends on the gate decomposition of the code. For two kinds of  decomposition, the fidelity was analytically calculated in Eqs. (\ref{eq:F_gateerror}}) to (\ref{eq:gate_result}). 
    \item{Repeat the error correction cycles only if you can reset the ancilla qubits. The optimal number of repetitions depends on the gate errors, as shown in Fig. \ref{fig:gate_error}}.
\end{itemize}
We envision that our
results will foster the design and interpretation of future
experiments on repetition code-based quantum memories. 

\begin{acknowledgments}{
We acknowledge helpful discussions and correspondence with J. K. Asbóth, A. Pályi, P. Boross, and Z. György.
This research was supported by the KDP-2021 program of the Ministry of Innovation and Technology from the source of the National Research, Development and Innovation Fund and the Ministry of Culture and Innovation, and the National Research, Development and Innovation Office within the Quantum Information National Laboratory of Hungary (Grant No. 2022-2.1.1-NL-2022-00004) and by
the NKFIH through the OTKA Grants FK 124723, FK 132146 and
FK 134437. }
\end{acknowledgments}

\bibliography{paper}
\end{document}